\documentclass[conference]{IEEEtran}

\usepackage{silence}
\usepackage{multicol}
\usepackage{multirow}
\usepackage{amssymb}
\usepackage{amsmath}
\usepackage{cases}
\usepackage{cancel}
\usepackage[T1]{fontenc}
\usepackage[utf8]{inputenc}
\usepackage[usenames, dvipsnames]{color}
\usepackage{pifont}
\usepackage{mathtools}
\usepackage{textcomp}
\usepackage{graphicx}
\usepackage{units}
\usepackage{subcaption}
\usepackage{amsthm}
\usepackage[dvipsnames]{xcolor}
\usepackage[compress]{cite}
\usepackage[english]{babel}
\usepackage{array}
\usepackage{setspace}
\usepackage[ruled,vlined,commentsnumbered]{algorithm2e}
\usepackage{bbm}

\newcommand{\Rhat}{\widehat{{R}}}
\newcommand{\Xhat}{\widehat{{X}}}
\newcommand{\Xtilde}{\Tilde{{X}}}

\newcommand{\Fn}{\mathcal{F}_n}
\newcommand{\Fm}{\mathcal{F}_m}
\newcommand{\Finvn}{\mathcal{F}_n^{-1}}
\newcommand{\Finvm}{\mathcal{F}_m^{-1}}

\newcolumntype{M}[1]{>{\centering\arraybackslash}m{#1}}

\title{Empowering 5G PRS-Based ISAC with Compressed Sensing}

\author{\IEEEauthorblockN{Esen Özbay, Pradyumna Kumar Bishoyi, and Marina Petrova}
\IEEEauthorblockA{Mobile Communications and Computing (MCC), RWTH Aachen University, Germany \\
Email: \{ozbay, pradyumna.bishoyi, petrova\}@mcc.rwth-aachen.de}}

\begin{document}

\maketitle

\begin{abstract}
To enable widespread use of Integrated Sensing and Communication (ISAC) in future communication systems, an important requirement is the ease of integration. A possible way to achieve this is to use existing communication reference signals for sensing, such as the 5G Positioning Reference Signal (PRS). Existing works have demonstrated promising results by using the PRS with classical signal processing techniques. However, this approach suffers from a loss of SNR due to the sparse resource allocation. In this work, we improve upon existing results by combining the 5G PRS with compressed sensing methods. We demonstrate that our method achieves better noise robustness compared to the existing works and has super-resolution properties, making it an ideal choice for range-Doppler map generation and target detection even in noisy environments.
\end{abstract}


\section{Introduction} \label{sec:Introduction}
Integrated sensing and communication (ISAC) has become one of the most promising technologies of the upcoming sixth-generation (6G) wireless systems for accommodating the diverse requirements of advanced services like autonomous driving, digital twins, smart factories, and extended reality (XR) ~\cite{NokiaSurvey5G6G}. The ISAC technology empowers the existing cellular base stations (BSs) with sensing capabilities, allowing the cellular wireless networks to not only provide high communication data rate but also accurate sensing and precise positioning services. The ISAC-enabled BS exploits the mutual benefit between sensing and communication~\cite{Zhang2022_survey41}. On the one hand, the sensing functionality offers assistance to communications in terms of beam training and beam tracking, i.e. sensing-assisted communication. On the other hand, communication-assisted sensing, where the existing communication signals are reused for sensing to gather prior information about the surrounding targets. The primary focus of our work is to study and design communication-assisted sensing by leveraging the existing 5G communication signals and making minimal modifications to the communication infrastructure, which remains an interesting open research problem.

In this direction, there are a few recent studies that investigate the feasibility and suitability of using current 5G New Radio (NR) orthogonal frequency division multiplexing (OFDM)-based communication signal for sensing purposes \cite{liu2022networked,cui_icassp,5G_PRS,csi-rs}. For example, Liu \textit{et al.} in \cite{liu2022networked} demonstrates that pilot signals possess unique benefits over data signals, primarily due to their strong auto-correlation characteristics. This makes them particularly suitable for sensing applications. Following this, authors in \cite{csi-rs} investigated the sensing performance of two 5G standard compliant downlink pilot signals, i.e., the channel state information reference signal (CSI-RS) and the demodulation reference signal (DMRS). The numerical simulation demonstrates that in regions with high signal-to-noise ratio (SNR), the CSI-RS pattern offers greater accuracy in range estimation compared to DMRS. However, for velocity estimation, both patterns yield the same level of accuracy. Further, in \cite{5G_PRS}, the authors conducted an analysis on the sensing performance of the positioning reference signal (PRS). They demonstrated that the PRS signal is both feasible and effective, especially when compared to DMRS and CSI-RS pilot signals. The PRS offers the following advantages: (i) its 31-bit long Gold sequence provides good auto-correlation property. (ii) There are four types of comb structures which allow for flexible time-frequency resource mapping. (iii) The different comb structure enables interference-free BS multiplexing, allowing multiple BSs to perform their sensing operation simultaneously. This motivates us to analyze the PRS-based sensing for ISAC system.

One of the main challenges for sensing parameter extraction using PRS is due to the sparsity in both the time and frequency domains. Applying the conventional 2D fast Fourier transform (FFT), i.e., the periodogram, to process the echo signal received from the sparse PRS structure results in a decrease in SNR and an increase in range-Doppler ambiguities. A suitable signal-processing tool to overcome this limitation is compressed sensing (CS), which can perform the estimation of sparse signals from under-sampled measurements~\cite{Eldar2012_CS}. The performance improvement of CS compared to the periodogram is demonstrated in~\cite{Nuss2017}, where CS was used with a stepped carrier OFDM signal to improve the output SNR. One of the popular CS algorithms is approximate message passing (AMP), which is an iterative method for solving the $\ell_1$-minimization problem~\cite[Ch. 9]{Eldar2012_CS}. Further, in~\cite{Knill18}, a complex extension to AMP, namely CAMP, is proposed to improve the sensing performance of an automotive radar operating at 77 GHz. The authors show that the CAMP algorithm significantly improved the SNR of the range-Doppler map while having a complexity similar to that of the periodogram.

In this paper, we study the potential of CAMP~\cite{anitori} to enhance the performance of PRS-based sensing in the context of an ISAC system. We consider a ISAC system where the BS performs monostatic sensing by transmitting PRS signal and collecting the echo. In our analysis, we include all four PRS comb patterns from the 5G standard, namely 2, 4, 6, and 12. Increasing the comb size results in a sparser signal pattern in the time-frequency domain. In contrast to the periodogram-based estimation used in \cite{5G_PRS,csi-rs}, we apply the CS-based algorithm to analyze the sensing echo and produce a range-Doppler map, which is then used for target detection. Different from~\cite{Knill18}, we deploy CAMP in a system based on existing 5G infrastructure. 
The \textit{contributions} of this paper are as follows:
\begin{itemize}
    \item We analyze the performance of a 5G-based ISAC system utilizing PRS waveform as a sensing signal. We employ the CAMP algorithm~\cite{Knill18} to process the received echo signal. The approach effectively reduces the degradation of the SNR caused by the non-continuous time-frequency PRS comb structure. This aspect is crucial to maintaining good sensing performance in terms of target detection.
    \item Through simulations, we demonstrate that the CAMP-based scheme significantly improves the SNR of the range-Doppler map. It also outperforms the periodogram-based scheme in terms of accurately distinguishing targets that are in close proximity to each other.
\end{itemize}

\textit{Notation}: The following notation is used in this paper: lowercase letters \mbox{($s$, $\lambda_c$, ...)} represent scalars; uppercase letters \mbox{($S$, $Y$, ...)} represent two-dimensional signals; and $\Fn$, $\Fm$, represent the Discrete Fourier Transform in the subcarrier and OFDM symbol axes, respectively. Similarly, $\Finvn$ and $\Finvm$ represent the inverse Fourier transforms.

The rest of this paper is organized as follows: in Section \ref{sec:SystemModel}, we explain the system model. Section \ref{sec:proposedMethod} describes the proposed CAMP algorithm-based sensing. Section \ref{sec:PerformanceEvaluation} discusses performance results and Section \ref{sec:Conclusion} concludes this paper.

\section{System Model}\label{sec:SystemModel}
\begin{figure}
    \centering
    \includegraphics[width=0.46\textwidth]{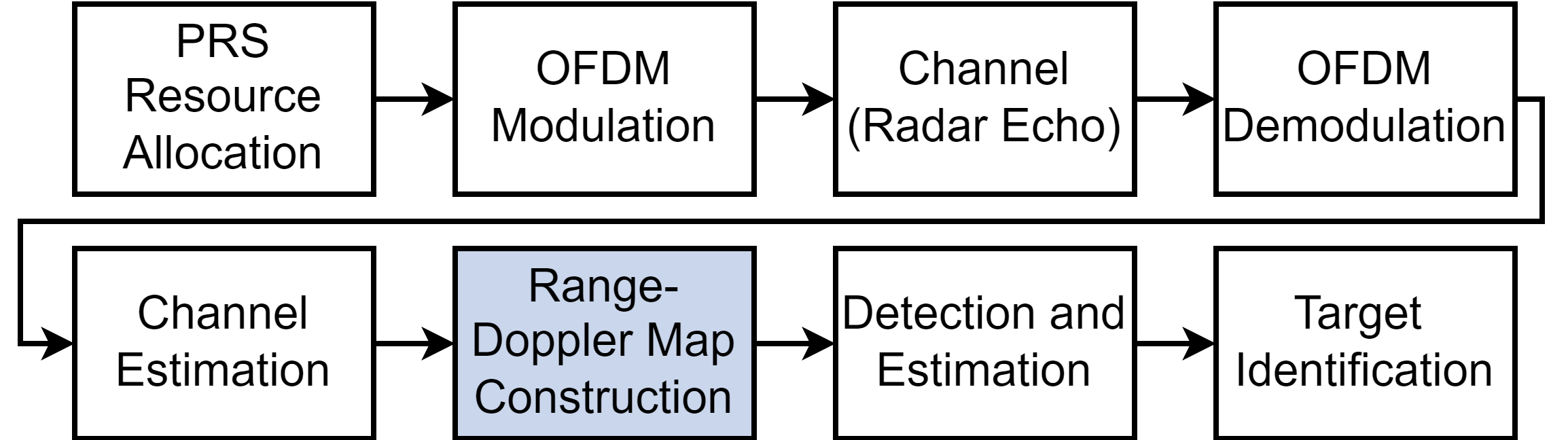}
    \caption{Block diagram overview of PRS-based ISAC system. This work focuses on the range-Doppler map construction.}
    \label{fig:blockdiagram}
\end{figure}
We consider a 5G-compliant BS operating in the millimeter-wave (mmWave) band, located in an urban environment. The BS acts as monostatic radar for target detection. We assume that the BS has full-duplex capability and can cancel out any self-interference at the receiver end. The BS transmits an OFDM grid consisting of $N_{sy}$ consecutive OFDM symbols with $N_{sc}$ subcarriers (SCs) each. The transmitted signal carries both PRS signal and data signal, which are scheduled orthogonal in time in order to reduce interference between them. The set of resource elements (REs)\footnote{RE is the smallest physical resource in 5G NR, corresponding to one subcarrier in frequency domain and one OFDM symbol in time domain} allocated for PRS is denoted as $\mathcal{P}$. The echo signal received from the target(s) is analyzed to create a range-Doppler map. Additionally, the map is utilized to detect the targets, enabling subsequent target identification. A block diagram of the transceiver module is given in Fig.~\ref{fig:blockdiagram}. Our primary objective is to effectively analyze the echo signal to provide a range-Doppler map that is free from any ambiguity. 
\subsection{The 5G PRS}
The PRS comb structure has four configurable parameters. The comb size, $K_c$, the time gap, $g$, the repetition factor, $F$, and the number of resource blocks, $N_{RB}$~\cite{Dahlman5G:Ch24}. The 5G standards define a \mbox{$K_c\times K_c$} `comb pattern for each value of $K_c$. The PRS resource allocation is obtained by repeating these comb patterns in the time-frequency grid. Note that the density of PRS symbols in the OFDM grid is equal to $1/K_c$.

\begin{figure}
    \centering
    \includegraphics[width=0.42\textwidth]{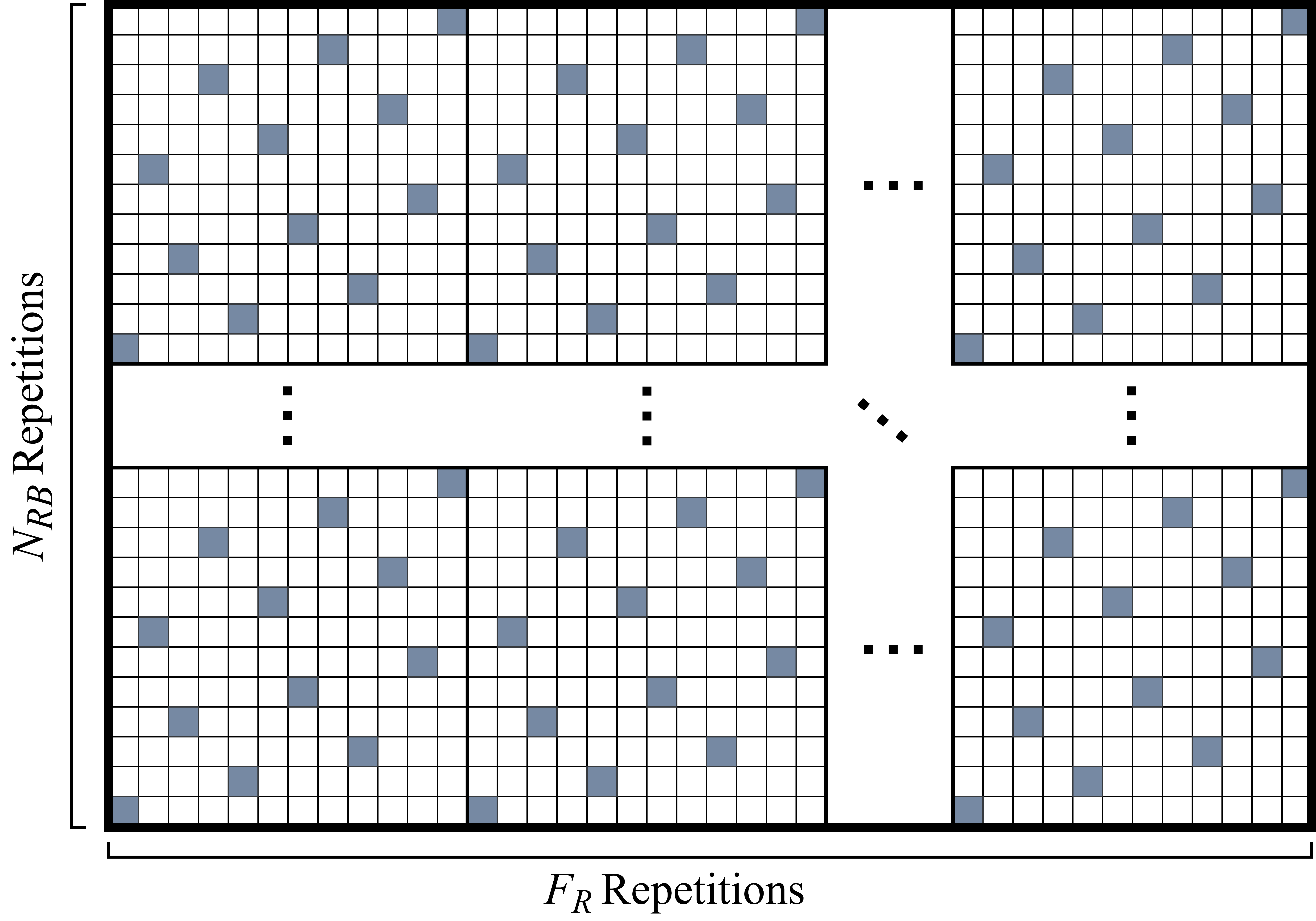}
    \caption{The resource allocation grid is obtained by repeating the comb patterns ($K_c=$12, $g=$1).}
    \label{fig:PRSrepetition}
\end{figure}

The total bandwidth spanned by the PRS is configured by the parameter $N_{RB}$. The comb pattern is repeated $12*N_{RB}/K_c$ times in the frequency axis in order to span $12*N_{RB}$ SCs. The number of OFDM symbols spanned by the PRS is configured by $g$ and $F$. The comb pattern is repeated a total of $F$ times in the time axis, with $K_c(g-1)$ symbols left blank between each repetition. Assuming that $g=1$ (i.e., the comb patterns are placed consecutively), the PRS signal spans a total of $K_cF$ OFDM symbols. A sample OFDM grid for $K_c=12$, $g=1$ is depicted in Fig. \ref{fig:PRSrepetition}.

\subsection{The Transmitted Signal}\label{Sec:txSig}
The signal transmitted by the BS can be represented by
\begin{equation}
        {S}     = \begin{bmatrix}
                        s_{0,0} 		& \hdots 	    & s_{0,(N_{sy}-1)} 	\\
                        \vdots		    & \ddots		& \vdots			\\
                        s_{N_{sc},0} 	& \hdots		& s_{(N_{sc}-1),(N_{sy}-1)} 	\\
                          \end{bmatrix},
\end{equation}
where $s_{n,m}$ represents a QPSK symbol carried on the $n$-th subcarrier of the $m$-th OFDM symbol. The corresponding time-domain transmitted signal is
\begin{equation}  
    s(t)=\sum_{m=0}^{N_{sy}-1}\sum_{n=0}^{N_{sc}-1}s_{n,m}e^{-j2\pi n\Delta ft}g(t-mT_s)\text{,}
\end{equation}
where $\Delta f$ is the OFDM subcarrier spacing (SCS), $g(t)$ is the pulse shape, $T_s=\frac{1}{\Delta f}+T_{CP}$ is the OFDM symbol duration including the cyclic prefix (CP), and $T_{CP}$ is the CP duration. Note that, as mentioned above, only the REs with $(n,m)\in\mathcal{P}$ are allotted for PRS. 

\subsection{The Received Echo Signal}
The channel experienced by the transmitted signal is the sum of echoes caused by the targets and the clutter. We model each target as a group of reflection centers~\cite{Buehren06}, making up a total of $L$ reflection centers. A reflection center with radar cross-section (RCS) $\sigma_l$, distance $R_l$ from the BS and radial speed $v_l$ causes a radar echo with delay $\tau_l$, Doppler shift $f_{D,l}$ and attenuation $\alpha_l$. These are given by
\begin{equation}\label{eq:alphatauf}
    \tau_l=\dfrac{2R_l}{c}, \quad f_{D,l}= \frac{2v_l}{\lambda_{c}}  \quad
    \alpha_l=\sqrt{\frac{G_t G_r \lambda_{c}^{2} \sigma_l}{(4\pi)^3 R_l^4}}\text{,}
\end{equation}
where $c$ is the speed of light, $\lambda_c$ is the carrier wavelength, and $G_t\text{,}\ G_r$ are the BS transmit and receive antenna gains, respectively. Note that only the line-of-sight (LoS) path between the BS and the target is considered in this analysis. Any non-line-of-sight (NLoS) path caused by the targets is assumed to be negligible. After OFDM demodulation, the effective channel caused by target $l$ over the $(n,m)$-th RE is~\cite[Sec. 3.2]{BraunThesis}
\begin{equation}
    h^{(l)}_{n,m} \triangleq \alpha_l e^{-j2\pi f_c\tau_l}e^{-j2\pi n \Delta f \tau_l}e^{j2\pi mT_sf_{D,l}}.
\end{equation}

Then, the received signal $y_{n,m}$ can be expressed as
\begin{equation}\label{eq:ReceivedSignal}
    \begin{split}
        y_{n,m}  &= s_{n,m}\bigg(h^{cl}_{n,m} + \sum^{L}_{l=1} h^{(l)}_{n,m}\bigg) + w_{n,m} \\
           &= s_{n,m}h_{n,m} + w_{n,m},
    \end{split}
\end{equation} 
where $w_{n,m}\sim\mathcal{CN}(0, N_0)$ is AWGN, and $h^{cl}_{n,m}$ is the channel created by the clutter.
\subsection{Channel Estimation and Range-Doppler Estimation}
In order to construct a range-Doppler map from the received echo signal $y_{n,m}$, $\forall (n,m)\in\mathcal{P}$, the BS must first estimate the effective channel over each OFDM symbol, $h_{n,m}$. The expression of the estimated channel, $\hat{h}_{n,m}$ is  
\begin{equation}\label{chan_est}
    \hat{h}_{n,m} = 
    \begin{cases}
        y_{n,m}/s_{n,m}, & (n,m)\in\mathcal{P} \\
        0,               & \text{otherwise.}
    \end{cases}
\end{equation}

Considering channel estimate in \eqref{chan_est}, typically, a periodogram is computed to generate the range-Doppler map. The output after performing 2D-FFT over the received signal $Y$ is 
\begin{equation}
    \Rhat = |\Xhat|^2 = |\Fm\{\Finvn\{H\}\}|^2.
\end{equation}
One of the major issues with applying the periodogram over PRS-based sensing is that PRS's sparse time-frequency structure results in a lower SNR gain. This leads to artefacts in the range-Doppler map and eventually increases the target misdetection rate. We illustrated this effect in our simulation results in Section \ref{sec:PerformanceEvaluation}. In order to address this issue, we employ CS-based processing, which has superior anti-noise capabilities and also can accurately estimate the sensing parameters from the sparse time-frequency structures.

\section{CS-based Processing over 5G PRS}\label{sec:proposedMethod}
In this section, we present the 2D-CAMP Algorithm, which is used to obtain the range-Doppler map. The 2D-CAMP is an iterative solution to the $\ell_1$-minimization problem, and yields a sparse estimate of the range-Doppler map from an incomplete OFDM grid~\cite{Knill18}. 

The 2D-CAMP algorithm is designed to yield a sparse output. In other words, most elements of the output of the 2D-CAMP are equal to zero. This makes the 2D-CAMP particularly fitted for sensing applications in sparse channels, such as the mmWave channel. Another property of the 2D-CAMP is that it is designed to be used with incomplete measurements. This makes 2D-CAMP suited for ISAC scenarios where there are no contiguous blocks of time-frequency resources that can be exclusively allocated to sensing, such as the case of PRS-based sensing. The 2D-CAMP is described in Algorithm \ref{alg:2DCAMP}.
\begin{algorithm} [t]
\setstretch{1.35}
\caption{The 2D-CAMP Algorithm~\cite{Knill18}}\label{alg:2DCAMP}
\KwIn{${H}$, $\mathcal{P}$, $N_{iter}$, $\tau_{CAMP}$}
\KwOut{$\Xhat$}
 Initialize $t = 1$, ${X}_0 = {0}_{N_{sc}\times N_{sy}}$, $E_0 = H$ \\
\While{$t < N_{iter}$}{
    $\Xtilde_t = \Fm\big\{\Finvn\big\{E_{t-1}\big\}\big\} + \Xhat_{t-1}$ \\
    $\sigma_t = \text{median}(|\Xtilde_t|)/\sqrt{2}$ \\
    $\Xhat_t=\text{soft}(\Xtilde_t, \tau_{CAMP}\sigma_t)$ \\
    $E_t = H - \mathbbm{1}_{(n,m)\in\mathcal{P}} \cdot \Finvm\big\{\Fn\big\{\Xhat_{t-1}\big\}\big\} + \frac{||\Xhat_t||_0} {N_{sc}N_{sy}}E_{t-1}$ \\
    $t = t + 1$ \\
}
$\Xhat = \Xhat_{N_{iter}}$
\end{algorithm}
\subsection{Description of the 2D-CAMP}
The inputs to the CAMP algorithm are as follows: the two-dimensional channel estimate signal $\hat{H}_{N_{sc}\times N_{sy}}$, number of iterations $N_{iter}$, the tunable thresholding parameter $\tau_{CAMP}$, and the PRS resource allocation set $\mathcal{P}$.

At each iteration $t$, the 2D-CAMP algorithm has the following steps:
\begin{enumerate}
    \item Calculate a noisy estimate $\Xtilde_t$ from the previous residual $E_{t-1}$,
    \item Calculate the noise variance $\sigma_t$ as the median of $\Xtilde_t$,
    \item From $\Xtilde_t$, obtain a sparse estimate $\Xhat_t$ by performing element-wise soft thresholding, where \mbox{$\lambda_{th}=\tau_{CAMP}\sigma_t$}, 
    \begin{equation}
        \text{soft}(x, \lambda_{th}) = \begin{cases}
            x\frac{|x|-\lambda_{th}}{|x|}, & |x| > \lambda_{th} \\
            0, & |x| \leq \lambda_{th},
        \end{cases}
    \end{equation}
    \item Calculate a new residual from the current estimate. 
\end{enumerate}
\subsection{Selection of the Parameter $\tau_{CAMP}$}
$\tau_{CAMP}$ is the only tunable parameter of the 2D-CAMP and determines the sensitivity of the algorithm. For higher values of $\tau_{CAMP}$, the soft thresholding step is more selective, leading to better noise robustness but a higher chance of missing weak targets. Lower values of $\tau_{CAMP}$ are more sensitive to both noise and to weak targets~\cite{anitori}. To achieve a desired false alarm probability $P_{FA}$, $\tau_{CAMP}$ can be selected as
\begin{equation}
    \tau_{CAMP} = -\sqrt{\ln(P_{FA})}.
\end{equation}

\subsection{Computational Complexity}
The 2D-CAMP algorithm contains a double FFT at each iteration. This is the most complex operation in the algorithm, therefore complexity of the 2D-CAMP is determined by the complexity of the 2D-FFT, $\mathcal{O}(NlogN \cdot MlogM)$. This is the same as the complexity of the periodogram~\cite{Knill18}. 


\section{Performance Evaluation}\label{sec:PerformanceEvaluation}
\begin{figure}
    \centering
    \includegraphics[width=0.47\textwidth]{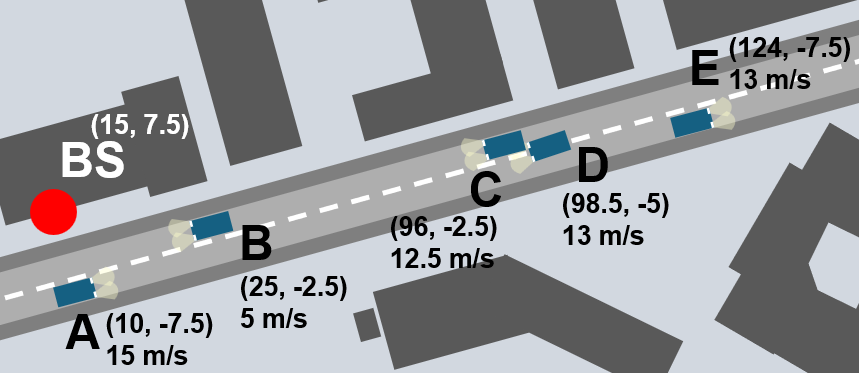}
    \caption{Simulation scenario. The position and the speed of each vehicle is given next to it.}
    \label{fig:scenario}
\end{figure}

\begin{table}
\caption{Waveform parameters.}
\centering
\begin{tabular}{| M{3.5cm} | M{1.25cm} |}
\hline
\textbf{Parameter}         &       \textbf{Value} \\
\hline
Carrier frequency, $f_c$        &       26 GHz \\
\hline
Subcarrier spacing, $\Delta f$  &       120 kHz \\
\hline
No. of subcarriers, $N_{sc}$    &       1620 \\
\hline
No. of symbols, $N_{sy}$        &       336  \\
\hline
Transmit power                  &       30 dBm \\
\hline
BS antenna gain                 &       18 dB \\
\hline
\end{tabular}
\label{tab:WaveformParams}
\end{table}
In this section, we evaluate the performance of CAMP algorithm on the 5G PRS-based sensing system. The simulation scenario is depicted in Fig.~\ref{fig:scenario}. We have considered a street named Kackertstrasse, which is located in Aachen, Germany. The BS is located at the beginning of the street and five vehicles are moving in lane. The 2D coordinates and the velocities of the BS and each vehicle are denoted in the Fig.~\ref{fig:scenario}. The vehicles were modelled using the reflection center model, where each vehicle is modelled as a group of reflection centers with different RCS values and positions~\cite{Buehren06}. Whether a particular reflection center is visible to the BS is determined according to the incidence angle. Each vehicle can have either one or two reflection centers that are visible to the BS. Further, we simulate the clutter with the MATLAB Ray Tracing tool by using a 3D-model of the street\footnote{The 3D street model was taken from openstreetmap.org.}.

The BS is operating in FR 2 band and transmits PRS signal configured with the parameters $K_c=12$, $N_{RB}=135$, $g=1$, $F=28$. This yields a total bandwidth of $200$ MHz and a CPI of $3$ ms. For the detection of targets (vehicles), the Constant False Alarm Rate (CFAR) method with $P_{FA}=10^{-7}$ is used. Note that the power levels in the range-Doppler maps are normalized so that the global maximum in each map is equal to 1, and all the figures depict the relative power. The rest of the system parameters are given in Table~\ref{tab:WaveformParams}.

\subsubsection{De-Noising}
\begin{figure}
    \vspace{0.1cm}
    \centering
    \includegraphics[width=0.485\textwidth]{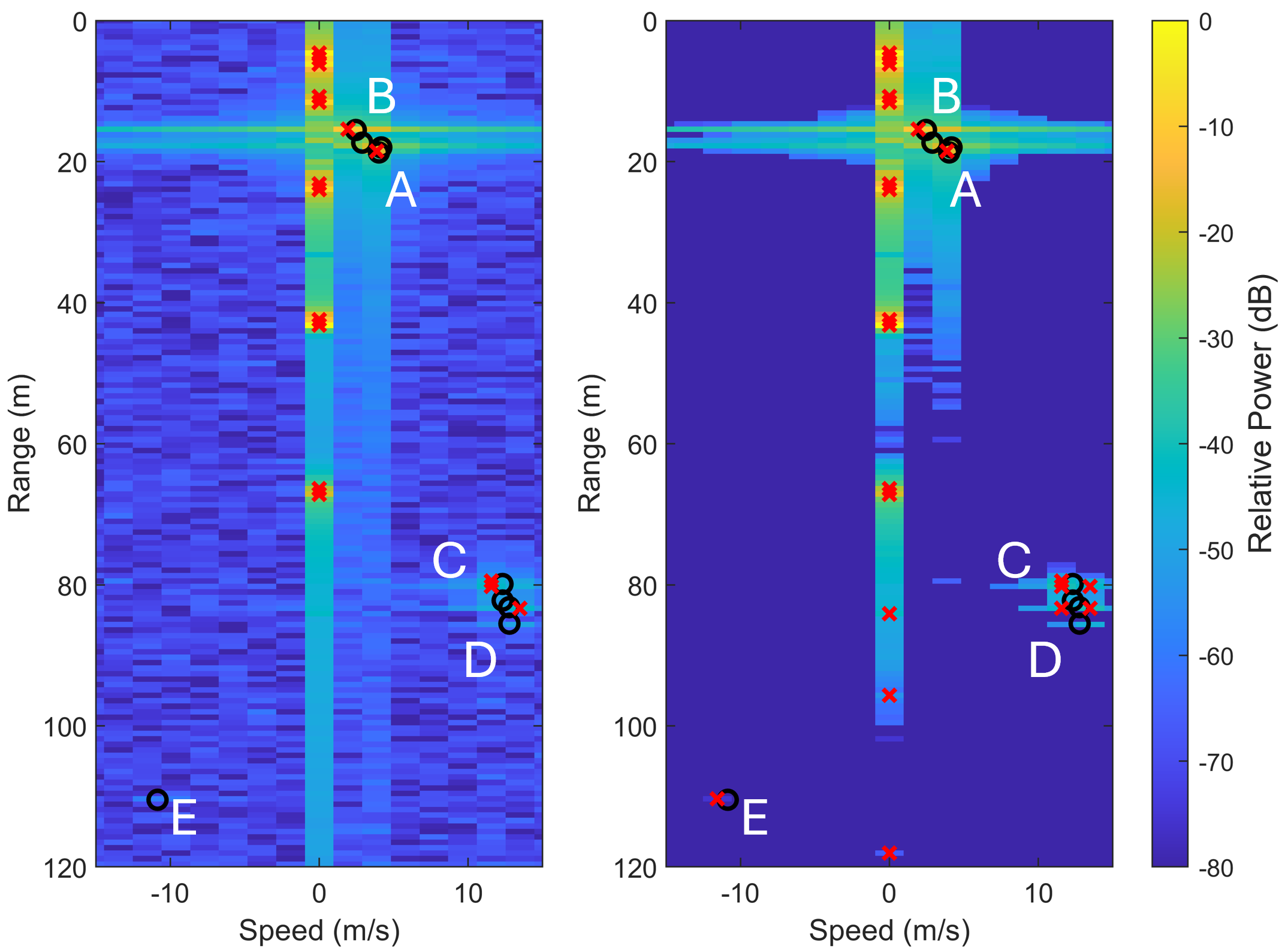}
    \caption{Range-Doppler maps obtained with the 2D FFT (left) and the 2D-CAMP (right). True values are marked with black circles, and CFAR detections are marked with red crosses. }
    \label{fig:rD_and_CFAR}
    \vspace{-0.4cm}
\end{figure}
Fig. \ref{fig:rD_and_CFAR} depicts the range-Doppler maps obtained with the 2D-FFT (the periodogram) and the 2D-CAMP ($\tau_{CAMP}=3.4$), respectively. Brighter colors represent higher received power for a given pair of range and speed values. Comparing the two figures, we observe that the 2D-CAMP yields a noiseless range-Doppler estimate of the targets. This is because the 2D-CAMP always produces sparse estimates, as opposed to the periodogram, which makes no assumptions about the map. In addition, we observe that the power level received from weaker targets is higher in 2D-CAMP when compared to the periodogram. For example, the relative power for target C increases from -46.7 dB to -43.2 dB in 2D-CAMP, which brings it above the detection threshold of CFAR. Furthermore, when comparing the two schemes, we can see that the 2D-CAMP improves the power level while maintaining a lower noise level and achieves higher SNR. This results in the generation of a noiseless range-Doppler map, leading to improved target detection. In Fig. \ref{fig:rD_and_CFAR}, the CFAR with the periodogram output can only detect four of the nine reflection centers that are present. In contrast, the CFAR with the 2D-CAMP output can detect seven of them.
\subsubsection{Super-Resolution Capabilities}
\begin{figure}
    \centering
    \includegraphics[width=0.485\textwidth]{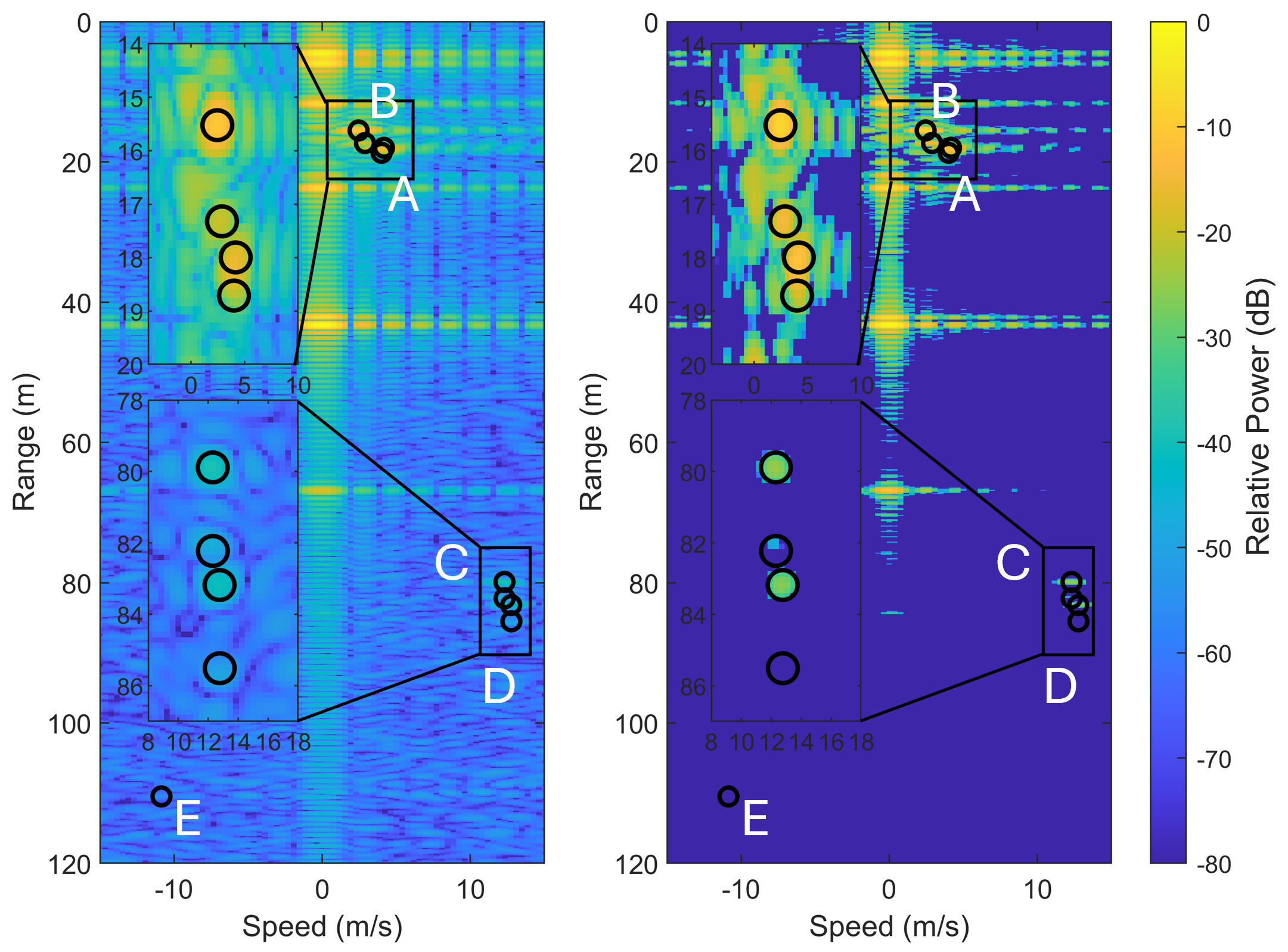}
    \caption{The range-Doppler maps for 2D-FFT (left) and 2D-CAMP (right), $N=5N_{sc}$, $M=5N_{sy}$.}
    \label{fig:highRes}
\end{figure}
In Fig.~\ref{fig:highRes}, we assess the super-resolution capabilities by increasing the FFT sizes of the periodogram. Specifically, we set $N=5N_{sc}$ and $M=5N_{sy}$, and compare the resulting output with that of the 2D-CAMP ($\tau_{CAMP}=4$) scheme. It can be seen that the 2D-CAMP produces a range-Doppler map that exhibits improved distinction between target peaks in comparison to the periodogram. For example, targets C and D produce two clearly distinguishable clusters when processed with the 2D-CAMP, whereas the periodogram produces peaks that are not clearly distinguishable. This is because with each iteration of 2D-CAMP, the thresholding step eliminates energy between the peaks and produces two distinct clusters. Conversely, the map obtained with the periodogram yields peaks that intersect with each other, making it difficult to distinguish close-by targets. This clearly shows that the super-resolution capability of the 2D-CAMP technique makes it well-suited for high-precision localization applications.
\subsubsection{The Effect of the Comb Size}
\begin{figure}
    \centering
    \includegraphics[width=0.45\textwidth]{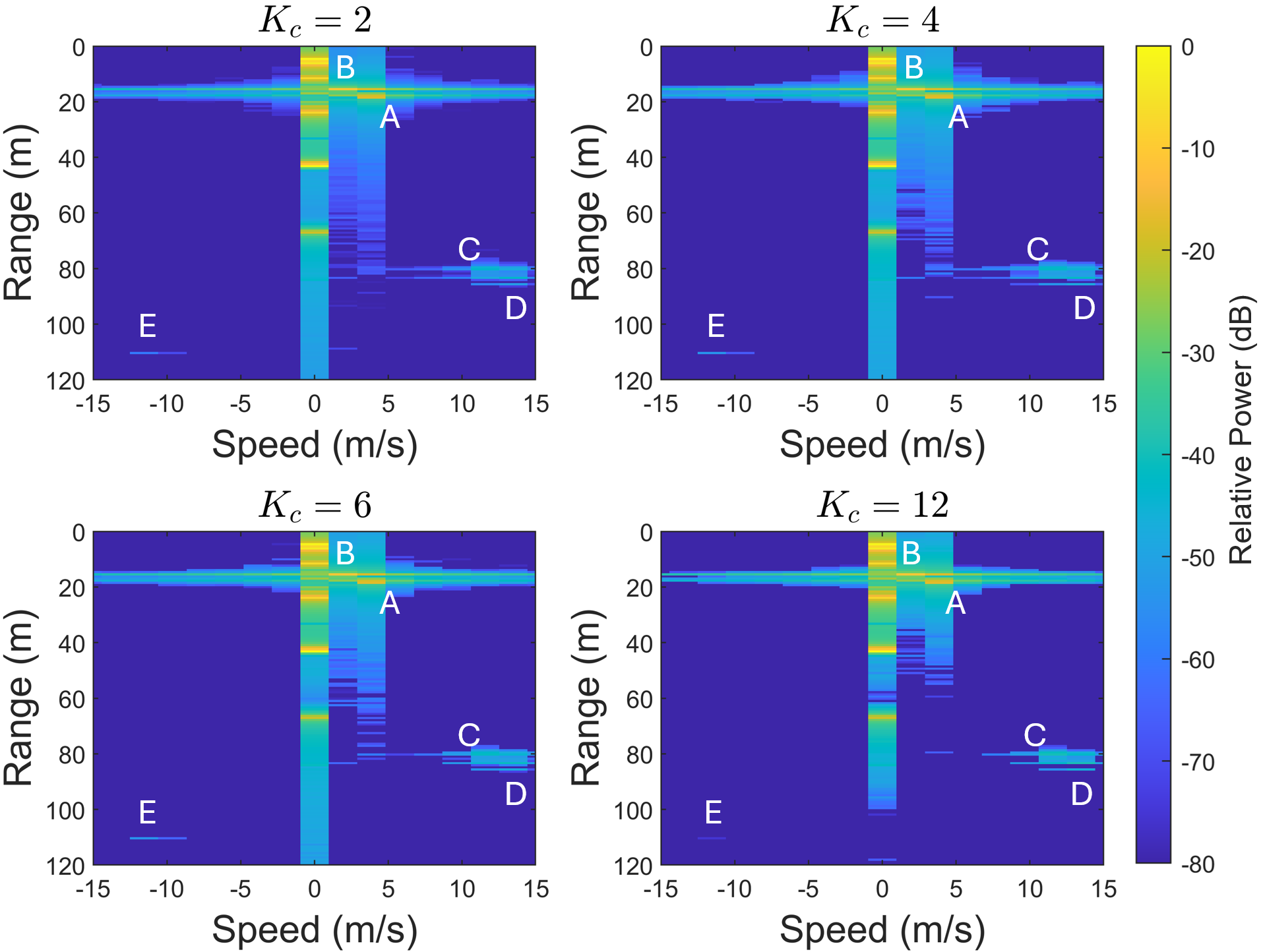}
    \caption{The range-Doppler map for $K_c = 2$, $4$, $6$, $12$.}
    \label{fig:combsizes}
\end{figure}

To fully understand the effect of time-frequency sparsity, in Fig.~\ref{fig:combsizes}, we compare the range-Doppler map obtained with the 2D-CAMP ($\tau_{CAMP}=3.4$) for different $K_c$ values. It can be observed that the range-Doppler maps are very similar for strong targets (A, B, C and D), even though constructed from signals with varying SNR values. However, as the value of $K_c$ increases, the received signal from the weaker target (E) becomes less distinct from the noise level. Adjusting the value $\tau_{CAMP}$ can effectively compensate for this issue, as it enhances the sensitivity of the 2D-CAMP to targets. Nevertheless, it is important to acknowledge that lowering $\tau_{CAMP}$ too much can cause the appearance of ghost targets. 

\section{Conclusion}\label{sec:Conclusion}
In this paper, we studied the 2D-CAMP, a CS algorithm, to obtain the range-Doppler map by using the 5G PRS as a sensing signal. We observed that the 2D-CAMP yielded a less noisy range-Doppler map than the periodogram method. Furthermore, 2D-CAMP showed a better capability to distinguish targets that are close to each other, while having the same asymptotic complexity as the periodogram. Given the limited resources available for sensing in ISAC scenarios, 2D-CAMP is a promising signal processing scheme for 5G-based sensing. In the future, we will explore different clutter suppression methods in combination with the CAMP to mitigate the effects of multipath.

\footnotesize
\section*{Acknowledgment}
This work was partially funded under the Excellence Strategy of the Federal Government and the Länder under grant "RWTH-startupPD 441-23".

\bibliographystyle{IEEEtran}
\bibliography{prs_based_frame.bib}

\begin{thebibliography}{10}
\providecommand{\url}[1]{#1}
\csname url@samestyle\endcsname
\providecommand{\newblock}{\relax}
\providecommand{\bibinfo}[2]{#2}
\providecommand{\BIBentrySTDinterwordspacing}{\spaceskip=0pt\relax}
\providecommand{\BIBentryALTinterwordstretchfactor}{4}
\providecommand{\BIBentryALTinterwordspacing}{\spaceskip=\fontdimen2\font plus
\BIBentryALTinterwordstretchfactor\fontdimen3\font minus \fontdimen4\font\relax}
\providecommand{\BIBforeignlanguage}[2]{{%
\expandafter\ifx\csname l@#1\endcsname\relax
\typeout{** WARNING: IEEEtran.bst: No hyphenation pattern has been}%
\typeout{** loaded for the language `#1'. Using the pattern for}%
\typeout{** the default language instead.}%
\else
\language=\csname l@#1\endcsname
\fi
#2}}
\providecommand{\BIBdecl}{\relax}
\BIBdecl

\bibitem{NokiaSurvey5G6G}
{T. Wild \textit{et al.}}, ``{Joint Design of Communication and Sensing for Beyond 5G and 6G Systems},'' \emph{{IEEE Access}}, vol.~9, pp. 30\,845--30\,857, 2021.

\bibitem{Zhang2022_survey41}
{J. A. Zhang \textit{et al.}}, ``{Enabling Joint Communication and Radar Sensing in Mobile Networks—A Survey},'' \emph{IEEE Commun. Surveys Tuts.}, vol.~24, no.~1, pp. 306--345, 2022.

\bibitem{liu2022networked}
{L. Liu \textit{et al.}}, ``{Networked Sensing in 6G Cellular Networks: Opportunities and Challenges},'' \emph{arXiv preprint arXiv:2206.00493}, 2022.

\bibitem{cui_icassp}
Y.~Cui, X.~Jing, and J.~Mu, ``{Integrated Sensing and Communications Via 5G NR Waveform: Performance Analysis},'' in \emph{Proc. IEEE ICASSP}, 2022, pp. 8747--8751.

\bibitem{5G_PRS}
{Z. Wei \textit{et al.}}, ``{5G PRS-Based Sensing: A Sensing Reference Signal Approach for Joint Sensing and Communication System},'' \emph{{IEEE Trans. Veh. Technol.}}, vol.~72, no.~3, pp. 3250--3263, 2023.

\bibitem{csi-rs}
{L. Ma \textit{et al.}}, ``{A Downlink Pilot Based Signal Processing Method for Integrated Sensing and Communication Towards 6G},'' in \emph{Proc. IEEE VTC2022-Spring}, 2022, pp. 1--5.

\bibitem{Eldar2012_CS}
Y.~C. Eldar and G.~Kutyniok, \emph{{Compressed sensing: Theory and Applications}}.\hskip 1em plus 0.5em minus 0.4em\relax Cambridge University Press, 2012.

\bibitem{Nuss2017}
B.~Nuss and T.~Zwick, ``{A novel interference mitigation technique for MIMO OFDM radar using compressed sensing},'' in \emph{Proc. IEEE EURAD}, 2017, pp. 98--101.

\bibitem{Knill18}
{C. Knill \textit{et al.}}, ``{High Range and Doppler Resolution by Application of Compressed Sensing Using Low Baseband Bandwidth OFDM Radar},'' \emph{IEEE Trans. Microw. Theory Techn.}, vol.~66, no.~7, pp. 3535--3546, 2018.

\bibitem{anitori}
{L. Anitori \textit{et al.}}, ``{Design and Analysis of Compressed Sensing Radar Detectors},'' \emph{IEEE Trans. Signal Process.}, vol.~61, no.~4, pp. 813--827, 2013.

\bibitem{Dahlman5G:Ch24}
E.~Dahlman, S.~Parkvall, and J.~Sköld, \emph{{5{G} {NR}: {T}he {N}ext {G}eneration {W}ireless {A}ccess {T}echnology}}, 2nd~ed.\hskip 1em plus 0.5em minus 0.4em\relax {Academic {P}ress}, 2021.

\bibitem{Buehren06}
M.~B{\"u}hren and B.~Yang, ``{Automotive Radar Target List Simulation Based on Reflection Center Representation of Objects},'' in \emph{Proc. WIT}, 2006, pp. 161--166.

\bibitem{BraunThesis}
K.~M. Braun, ``\BIBforeignlanguage{english}{{OFDM Radar Algorithms in Mobile Communication Networks}},'' Ph.D. dissertation, {KIT, Germany}, 2014.

\end{thebibliography}

\end{document}